\documentclass[aps,prl,showpacs,graphicx,twocolumn,,superscriptaddress,10pt]{revtex4-1}%
\usepackage{mathrsfs}
\usepackage{times}
\usepackage{amsmath}
\usepackage{graphicx}
\usepackage{sidecap}
\usepackage{txfonts}
\usepackage{color}
\usepackage[colorlinks=true,citecolor=blue,linkcolor=magenta]{hyperref}
\usepackage{subeqnarray}
\usepackage{verbatim}
\usepackage{epstopdf}
\usepackage{cases}
\usepackage[symbol]{footmisc}

\DefineFNsymbols{nu}{\dag *}\setfnsymbol{nu}

\begin{document}

\title{Quantifying Quantum Coherence in Experimentally-Observed Neutrino Oscillations}

\author{Xue-Ke Song}
\email{These authors contributed equally to this work.}

\affiliation{Shenzhen Institute for Quantum Science and Engineering and Department of Physics, Southern University of Science and Technology, Shenzhen 518055, China}
\affiliation{Department of Physics, Southeast University, Nanjing 211189, China}
\affiliation{Shenzhen Key Laboratory of Quantum Science and Engineering, Southern University of Science and Technology, Shenzhen, 518055, China}

\author{Yanqi Huang}
\email{These authors contributed equally to this work.}
\affiliation{School of Physics and State Key Laboratory of Nuclear Physics and
Technology, Peking University, Beijing 100871,
China }

\author{Jiajie Ling}
\affiliation{School of Physics, Sun Yat-Sen University, Guangzhou 510275, China}

\author{Man-Hong Yung}
\email{Corresponding author: yung@sustc.edu.cn}
\affiliation{Shenzhen Institute for Quantum Science and Engineering and Department of Physics, Southern University of Science and Technology, Shenzhen 518055, China}
\affiliation{Shenzhen Key Laboratory of Quantum Science and Engineering, Southern University of Science and Technology, Shenzhen, 518055, China}

\date{\today }

\begin{abstract}
Neutrino oscillation represents an intriguing physical phenomenon where the quantumness can be maintained and detected over a long distance. Previously, the non-classical character of neutrino oscillation was tested with the Leggett-Garg inequality, where a  clear violation of the classical bound was observed [J. A. Formaggio \emph{et al}., Phys. Rev. Lett. \textbf{117}, 050402 (2016)]. However, there are several limitations in testing neutrino oscillations with the Leggett-Garg inequality. In particular, the degree of violation of the Leggett-Garg inequality cannot be taken as a ``measure of quantumness". Here we focus on quantifying the quantumness of experimentally-observed neutrino oscillation, using the tools of recently-developed quantum resource theory. We analyzed ensembles of reactor and accelerator neutrinos at distinct energies from a variety of neutrino sources, including Daya Bay (0.5 km and 1.6 km), Kamland (180 km), MINOS (735 km), and T2K (295 km). The quantumness of the three-flavoured neutrino oscillation is characterized within a $3\sigma$ range relative to the theoretical prediction. It is found that the maximal coherence was observed in the neutrino source from the Kamland reactor. However, even though the survival probability of the Daya Bay experiment did not vary significantly (dropped about 10 percent), the coherence recorded can reach up to 40 percent of the maximal value. These results represent the longest distance over which quantumness were experimentally determined for quantum particles other than photons.
\end{abstract}

\maketitle


\emph{Introduction.---}  The phenomenon of neutrino oscillation was proposed for more than half a century~\cite{Pontecorvo:1957qd, Maki:1962mu}. Since then, compelling experimental evidences of the transitions between different neutrino flavors have been obtained from different sources, including solar~\cite{e1,e2}, atmosphere~\cite{e3}, reactor~\cite{e4,e5} and accelerator neutrinos~\cite{e6,e7,e8,e9}. In the three-generation neutrino framework, neutrinos and antineutrinos are produced simultaneously and detected in three different flavors, namely electron $e$, muon $\mu$, and tau $\tau$ leptons. The flavor states are linear combination of the mass states~\cite{Camilleri:2008zz,Duan:2010bg}. Neutrino oscillation implies that a given flavor may change into another flavor during the propagating, caused by the nonzero neutrino mass and neutrino mixing. Recently, a number of refined measurements and analyses on the  oscillations parameters have been presented~\cite{dayabaydata,kamlanddata,minosdata,t2kdata}. However, the justification of neutrino oscillation is based on a crucial assumption that the different neutrino states are {\it well coherent} during its propagating; this assumption of quantum coherence still needs to be verified carefully, as it leads to considerable constrains in ultra-high energy or astronomical scales~\cite{Bustamante:2015waa,Huang:2015flc}. Furthermore, to explore the possibility of utilizing neutrino oscillations for future applications in quantum information processing, an important step is to verify the ``quantumness" in neutrino oscillations.

The question is, how might one test the quantumness of neutrino oscillations? In recent years, the idea of testing neutrino oscillations using  Leggett-Garg inequality (LGI) \cite{Leggett,Emary} have been considered~\cite{Gangopadhyaytwo, Formaggio, Fu,Gangopadhyaythree}; it is suggested that experimentally-observed neutrino oscillations can violate the classical limits imposed by the LGI. However, there are several fundamental problems associated with testing neutrino oscillations with LGI. (i) the LGI was originally designed to test the concept of macroscopic realism for macroscopic objects; violation of a LGI means that the system maybe neither a macroscopic reality nor non-invasive
measurement can be performed on it. However, these two conditions are not strictly satisfied for elementary particles probed in the current experimental settings. (ii) Experimental violation of LGI assumed~\cite{Formaggio} that there are only two neutrino states, instead of three.
  (iii) Even if LGI is violated, one still cannot {\it quantify} the amount of coherence directly through the violation. The situation is similar to the case where Bell's inequalities cannot be utilized for quantifying quantum entanglement.

In the context of quantum information, coherence is a fundamental concept that can be rigorously characterized in the context of quantum resource theory~\cite{Baumgratz}. Similar to quantum entanglement, there are different measures for coherence (see Refs. \cite{ Streltsov, Hu} for a review). Among them, the $l_1$-norm of coherence,
\begin{equation}
\mathcal{C} (\rho)=\sum_{i\neq j} \ |\rho_{ij}|  \ \ge 0 \ ,
\end{equation}
is probably the most accessible one, for the neutrino experiments performed; it is equal to a summation over the absolute values of all the off-diagonal elements $\rho_{ij}$ of a given density matrix $\rho$. Moreover, it fulfills all the necessary requirements of a coherence measure:
non-negativity, monotonicity, strong monotonicity, and convexity. The maximal possible value of $\mathcal{C}$ is bounded by $\mathcal{C}_{\max}=d-1$ with $d$ being dimension of the density matrix $\rho$. In particular,  $\mathcal{C}_{\max}=2$ for three-flavored neutrino oscillations considered in this work.

There are many reasons for quantifying quantum coherence. For example, one can estimate the required copies for converting quantum states with different amount of coherence through incoherent operations, which is similar to entanglement distillation~\cite{winter}. Many results indicate that coherence can be regarded as resources for quantum algorithms~\cite{matera}, quantum~channel discrimination~\cite{piani}, and quantum thermodynamics~\cite{Streltsov,gour}.

Here, we present a method for quantifying the quantumness of neutrino oscillation, with the use of a coherence measure developed in quantum resource theory. Through analyzing the experimental data from different sites, including Daya Bay, Kamland, T2K, and MINOS, we  study the coherence in the dynamics of the three-flavor neutrino oscillations. We conclude that a significant amount of quantum coherence exists from all four sources of neutrinos. In particular, the Kamland collaboration recorded the highest value of coherence; many events are close to the theoretical maximal value of 2. Furthermore, from the Daya Bay data, even though the transition probabilities from electron neutrino to other flavors are less than $10\%$, the coherence can reach as much as $0.8$ ($40$ precent of the maximal value). Although currently utilizing neutrino oscillations for practical applications remains a major technological challenge, quantification of the quantumness of neutrino oscillation represents the first step towards this goal.


\emph{The neutrino model.---} In the three-generation framework, a neutrino oscillation involves the mixing between the flavor states $|\nu_e\rangle$, $|\nu_\mu\rangle$, $|\nu_\tau\rangle$ which are superpositions of the mass eigenstates $|\nu_1\rangle$, $|\nu_2\rangle$, $|\nu_3\rangle$
 (here, $(e, \mu, \tau)$ represents the neutrino flavor state and (1, 2, 3) labels the neutrino
mass state). The explicit relation is given by a $3\times 3$ unitary matrix $U$,
i.e., the Pontecorvo-Maki-Nakagawa-Sakata (PMNS) matrix \cite{Maki}.
Each flavor state is a linear superposition
of the mass eigenstates: $|\nu_{\alpha}\rangle=\sum_{k}U_{\alpha k}|\nu_{k} \rangle$, where $\alpha=e,\mu,\tau$ and $k=1,2,3$. In the standard parametrization, $U$ is characterized by three mixing angles $(\theta_{12}, \theta_{13}, \theta_{23})$ and a charge
coujugation and parity (CP) violating phase $\delta_{CP}$,
\begin{align}  
\left(
           \begin{array}{ccc}
             c_{12}c_{13} &  s_{12}c_{13} &  s_{13}e^{-i\delta_{CP}} \\
             -\!s_{12}c_{23}\!-\!c_{12}s_{13}s_{23}e^{i\delta_{CP}} & c_{12}c_{23}\!-\!s_{12}s_{13}s_{23}e^{i\delta_{CP}} & c_{13}s_{23}\\
             s_{12}s_{23}\!-\!c_{12}s_{13}c_{23}e^{i\delta_{CP}} & -\!c_{12}s_{23}\!-\!s_{12}s_{13}c_{23}e^{i\delta_{CP}} & c_{13}c_{23} \\
           \end{array}
         \right),
\end{align}
where $c_{ij}\equiv \cos\theta_{ij}$ and $s_{ij}\equiv \sin\theta_{ij}$ $(i,j=1,2,3)$. Currently, there are little conclusive evidences about the CP phase, we assume that it vanishes in the following discussion.

The massive neutrino states are eigenstates of the time-independent free
Dirac Hamiltonian $H$ with an energy $E_k$, and its time evolution satisfies the relativistic quantum mechanics dynamical equation.
Explicitly, during the neutrino propagation, the wave function solution is given by $|\nu_{k}(t)\rangle=e^{-iE_kt/\hbar}|\nu_{k}(0)\rangle$, which implies that the time evolution of flavor neutrino states is given by $|\nu_{\alpha}(t)\rangle=a_{\alpha e}(t)|\nu_{e}\rangle+a_{\alpha \mu}(t)|\nu_{\mu}\rangle+a_{\alpha \tau}(t)|\nu_{\tau}\rangle$, where $a_{\alpha \beta}(t)\equiv \sum_kU_{\alpha k}e^{-iE_kt/\hbar}U_{\beta k}^\ast$. Finally, the probability for detecting $\beta$ neutrino, given that the initial state is in the $\alpha$ neutrino state, is given by \cite{Garcia2},
\begin{eqnarray}  
P_{\alpha\beta}&=&\delta_{\alpha\beta}-4\sum_{k>l}\textrm{Re}(U_{\alpha k}^\ast U_{\beta k}U_{\alpha l}U_{\beta l}^\ast)\sin^2\left(\Delta m^2_{kl}\frac{Lc^3}{4\hbar E}\right)\nonumber\\
&+&2\sum_{k>l}\textrm{Im}(U_{\alpha k}^\ast U_{\beta k}U_{\alpha l}U_{\beta l}^\ast)\sin\left(\Delta m^2_{kl}\frac{Lc^3}{2\hbar E}\right),
\label{tp}
\end{eqnarray}
where $\Delta m^2_{kl}\equiv m^2_{k}-m^2_{l}$, $E$ is the energy of the neutrino which is different for different neutrino experiments,
and $L=ct$ (with $c$ being the speed of light) is the distance traveled by the neutrino particle. Note that in the neutrino experiments, one may vary the energy $E$, instead of time, for probing the variation of the transition probabilities (e.g. see Ref.~\cite{Formaggio}).

For analyzing the experimental data of neutrino oscillations, it is
convenient to write the oscillatory term of Eq. (\ref{tp}), $\sin^2\left(\Delta m^2_{kl}\frac{Lc^3}{4\hbar E}\right)$, in a simple form~\cite{Giunte},
\begin{align}  
\sin^2\left(\Delta m^2_{kl}\frac{Lc^3}{4\hbar E}\right)=\sin^2\left(1.27\Delta m^2_{kl}[eV^2]\frac{L[km]}{E[GeV]}\right).
\end{align}
Note that the oscillation probabilities depend on six independent parameters (three mixing angles, mass squared difference, distance and
energy) and four of them can be experimentally determined. For the following order of the neutrino mass spectrum, $m_{1}<m_{2}<m_{3}$, the best
fit values and the $3\sigma$ ranges of the three-flavor oscillation parameters are listed in Table \ref{tab:1}. Below, we shall consider separately the electron neutrino oscillations at Daya Bay (0.5 km and 1.6 km) and Kamland (180 km), and muon neutrino oscillations at MINOS (735 km) and T2K (295 km).

\begin{table}[]
      \caption{The neutrino mixing parameters in normal heirachy from the global fit results \cite{Garcia2}.}
      \label{tab:1}
      \begin{ruledtabular}
      \centering
      \begin{tabular}{ccc}

       parameter & best fit$\pm 1\sigma$\  & $3\sigma$ range\\
      \hline
      \noalign{\vspace{0.5ex}}
      $\Delta m^2_{21}/10^{-5}~{\rm eV^2}$ & $7.50_{-0.17}^{+0.19}$ & $7.02 \to 8.09$\\
      \noalign{\vspace{0.5ex}}
      $\Delta m^2_{31}/10^{-3}~{\rm eV^2}$ & $2.457_{-0.047}^{+0.047}$ & $2.317 \to 2.607$\\
      \noalign{\vspace{0.5ex}}
      $\theta_{12}/^\circ$ & $33.48_{-0.75}^{+0.78}$ & $31.29 \to 35.91$ \\
      \noalign{\vspace{0.5ex}}
      $\theta_{23}/^\circ$ & $42.3_{-1.6}^{+3.0}$ & $38.2 \to 53.3$ \\
      \noalign{\vspace{0.5ex}}
      $\theta_{13}/^\circ$ & $8.50_{-0.21}^{+0.20}$ & $7.85 \to 9.10$ \\

      \end{tabular}
      \end{ruledtabular}
\end{table}

%
%
%
%

\begin{figure}[!h]
\begin{center}
\includegraphics[width=7.8 cm,angle=0]{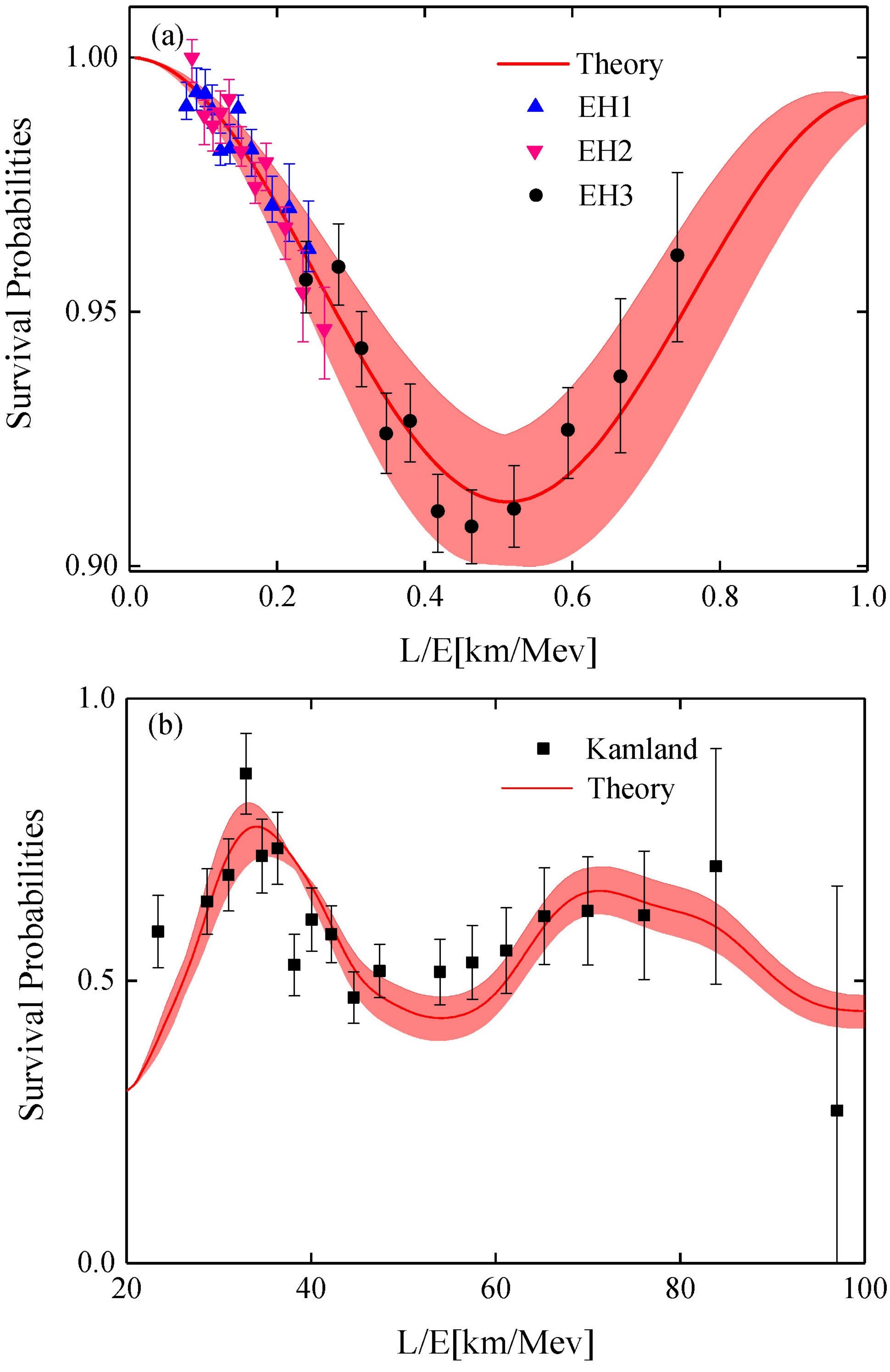}
\caption{ (a) Short range neutrino survival probabilities for the initial electron neutrino in theory (red, solid line) and the data of Daya Bay collaboration in three underground experimental halls
(EH1: blue, upper triangle, EH2: pink, lower triangle, EH3: black, circle) taken from Refs. \cite{dayabaydata}
(black, square) with ratio $L/E$ changing are plotted; (b) Long range neutrino survival probabilities for
the initial electron neutrino in theory (red, solid line) and  the data of Kamland collaboration taken from Ref. \cite{kamlanddata} (black, square) with ratio $L/E$ changing are shown.
The red band indicates a $3\sigma$ confidence interval around the
fitted prediction.}\label{fig1}
\end{center}
\end{figure}

\begin{figure}[!h]
\begin{center}
\includegraphics[width=7.8 cm,angle=0]{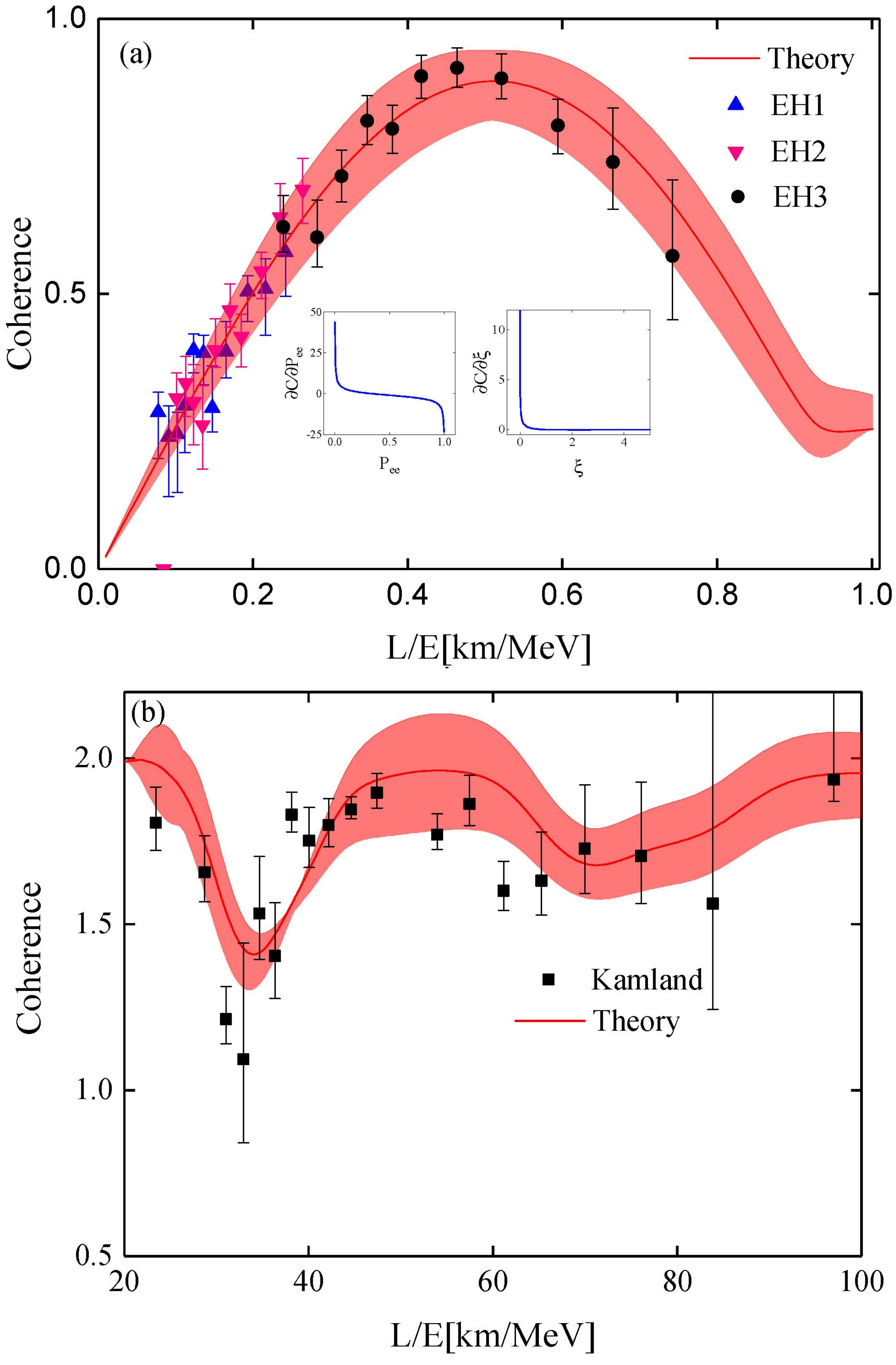}
\caption{The coherence in theory (red, solid line) with a $3\sigma$ confidence interval around the
fitted prediction (red band) and experiment (black, square) with error bar
in (a) Daya Bay (black, circle) and (b) Kamland collaborations (black, square)
for three-flavored neutrino oscillations as the function of ratio $L/E$ between
the traveled distance and neutrino energy are shown. The inset shows that the derivative of coherence
with respect to survival probability and the ratio $\xi$, respectively. Taking the error bar into consideration, the data are in consistent with the theoretic $3\sigma$ range in the short distance case.}\label{fig2}
\end{center}
\end{figure}


\emph{Coherence in electron antineutrino oscillations.---} If an electron neutrino is produced at the initial time $t=0$, then its time evolution is given by, $|\nu_{e}(t)\rangle=a_{ee}(t)|\nu_{e}\rangle+a_{e\mu}(t)|\nu_{\mu}\rangle+a_{e\tau}(t)|\nu_{\tau}\rangle$, where the probabilities for finding the neutrino in state $|\nu_{e}\rangle$,
$|\nu_{\mu}\rangle$ and $|\nu_{\tau}\rangle$ are, respectively,
$P_{ee}(t)=|a_{ee}(t)|^2$, $P_{e\mu}(t)=|a_{e\mu}(t)|^2$ and $P_{e\tau}(t)=|a_{e\tau}(t)|^2$.
In Fig. \ref{fig1}, we plot the variations of the survival probabilities in neutrino oscillations from Daya Bay~\cite{dayabaydata} and Kamland~\cite{kamlanddata}, as a function of $L/E$; these sites make use of $\beta$-decay
to produce a source of the electron antineutrino, but with different baselines and energy, changing the ratio of $L/E$. The Daya Bay Collaboration used the fully constructed Daya Bay Reactor Neutrino Experiment
as a new measurement of electron antineutrino disappearance, and it covers energy between 1 to 8 $MeV$,
which signifies a effective ratio $L/E$ in a range [0, 1] with dimension $km/MeV$. However, the Kamland Collaboration demonstrated the oscillatory nature of neutrino flavor transformation by observing electron antineutrinos
with energies of a few $MeV$ from nuclear reactors about 180 $km$ away, corresponding to the range [0, 100] in terms of the ratio $L/E$.

For the short-range oscillations (with small $L/E$) at Daya Bay, shown in Fig. \ref{fig1} (a), the survival probability of $\nu_{e}$ is always higher than 0.9; the probabilities of detecting the other favors are relatively small. It reaches the minimal point at around $L/E=0.5$. On the other hand, in Fig. \ref{fig1} (b), the long-range neutrino oscillations in Kamland involve significant contributions from all three flavors. Technically, the length of the baseline for the theoretical prediction of survival probability in Kamland collaboration
was taken as the average value, it thus presents a relatively smooth manner than a cosine one from the Eq. (\ref{tp}). It coincides with the data given by the Kamland experiment.

To quantify the coherence, we shall focus on the off-diagonal elements of the density matrix $\rho \left( t \right) = \left| {{\nu _e}\left( t \right)} \right\rangle \left\langle {{\nu _e}\left( t \right)} \right|$, in the basis $\{|\nu_{e}\rangle,|\nu_{\mu}\rangle,|\nu_{\tau}\rangle\}$,
\begin{align}  
\rho(t)=\left(
           \begin{array}{ccc}
             |a_{ee}(t)|^2 & a_{ee}(t)a_{e\mu}^*(t) & a_{ee}(t)a_{e\tau}^*(t) \\
             a_{ee}^*(t)a_{e\mu}(t) & |a_{e\mu}(t)|^2 & a_{e\mu}(t)a_{e\tau}^*(t) \\
             a_{ee}^*(t)a_{e\tau}(t) & a_{e\mu}^*(t)a_{e\tau}(t) & |a_{e\tau}(t)|^2 \\
           \end{array}
         \right) \ ,
\end{align}
where the coherence is given by $\mathcal{C}=2|a_{ee}(t)a_{e\mu}(t)|+2|a_{ee}(t)a_{e\tau}(t)|+2|a_{e\mu}(t)a_{e\tau}(t)|$. Equivalently, it can be expressed in terms of the transition probabilities, i.e.,
\begin{align}  
\mathcal{C}_e=2\left(\sqrt{P_{ee}(t)P_{e\mu}(t)}+\sqrt{P_{ee}(t)P_{e\tau}(t)}+\sqrt{P_{e\mu}(t)P_{e\tau}(t)}\right),
\label{coherence}
\end{align}
Note that transitions probabilities are subject to the normalization constraint: $\sum_{\alpha}P_{\alpha\beta}=\sum_{\beta}P_{\alpha\beta}=1$ ($\alpha,\beta=e,\mu,\tau$). A partial derivative of coherence with
respect to survival probability presents that a tiny change of $P_{ee}(t)$ will lead to a relatively drastic
variation of coherence when the survival probability takes a value either smaller than 0.1 or larger than 0.9, see the inset of Fig. \ref{fig2} (a). Experimentally, only the survival probability is given, to quantify the measured coherence, ratio $\xi=P_{e\tau}(t)/P_{e\mu}(t)$ is determined from the theoretical prediction; while the coherence changes gently with the ratio $\xi$ changing when $\xi>0.5$, there is a dramatic change of coherence for $\xi<0.5$.

The coherence in theory and experiment for three-flavored neutrino oscillations as a function of
ratio $L/E$ are plotted in Fig. \ref{fig2}. The experimental data exhibit a good agreement with the theoretical predictions. For the Daya Bay collaboration, the coherence may reach about $0.8$ at $L/E=0.5$, even though the transition probabilities to other flavors is less than $10 \%$.

On the other hand, the coherence from the neutrino oscillation at Kamland is in general higher (see Fig.~\ref{fig2} (b)), even reaches the maximum value of 2 for three-flavored
neutrino oscillations. It implies more quantum resource can be used in the long-distance propagating of neutrino
in the Kamland experiment.




\begin{figure}[t]
\begin{center}
\includegraphics[width=8 cm,angle=0]{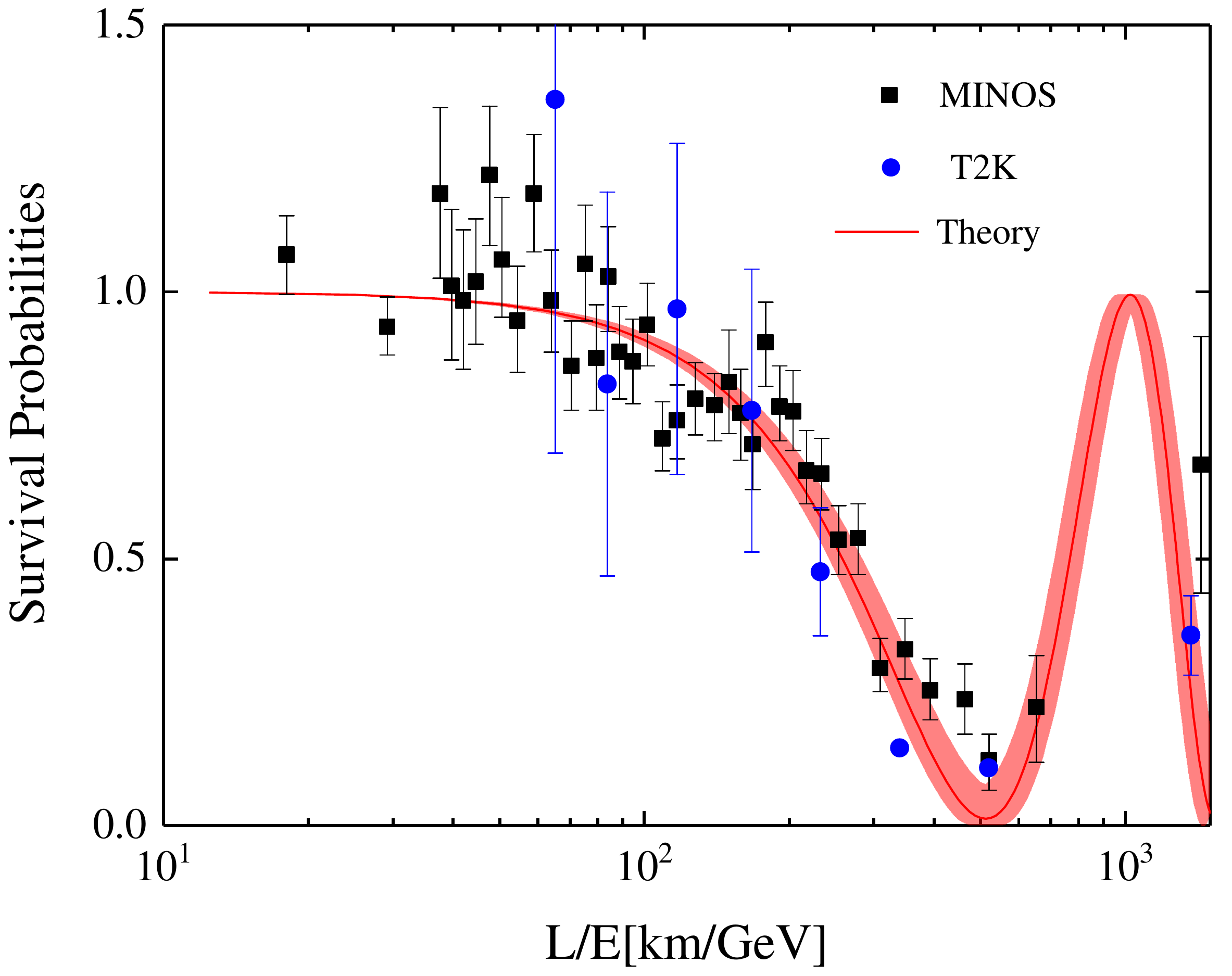}
\caption{Long range neutrino survival probabilities for an initial muon neutrino in theory (red, solid line) and the data
 of MINOS (black, square) and T2K (blue, circle) collaborations taken from Refs. \cite{minosdata}
and \cite{t2kdata} with ratio $L/E$ changing are shown, respectively. The red band indicates a $3\sigma$ confidence interval around the
fitted prediction.}\label{fig3}
\end{center}
\end{figure}

\begin{figure}[t!]
\begin{center}
\includegraphics[width=8cm,angle=0]{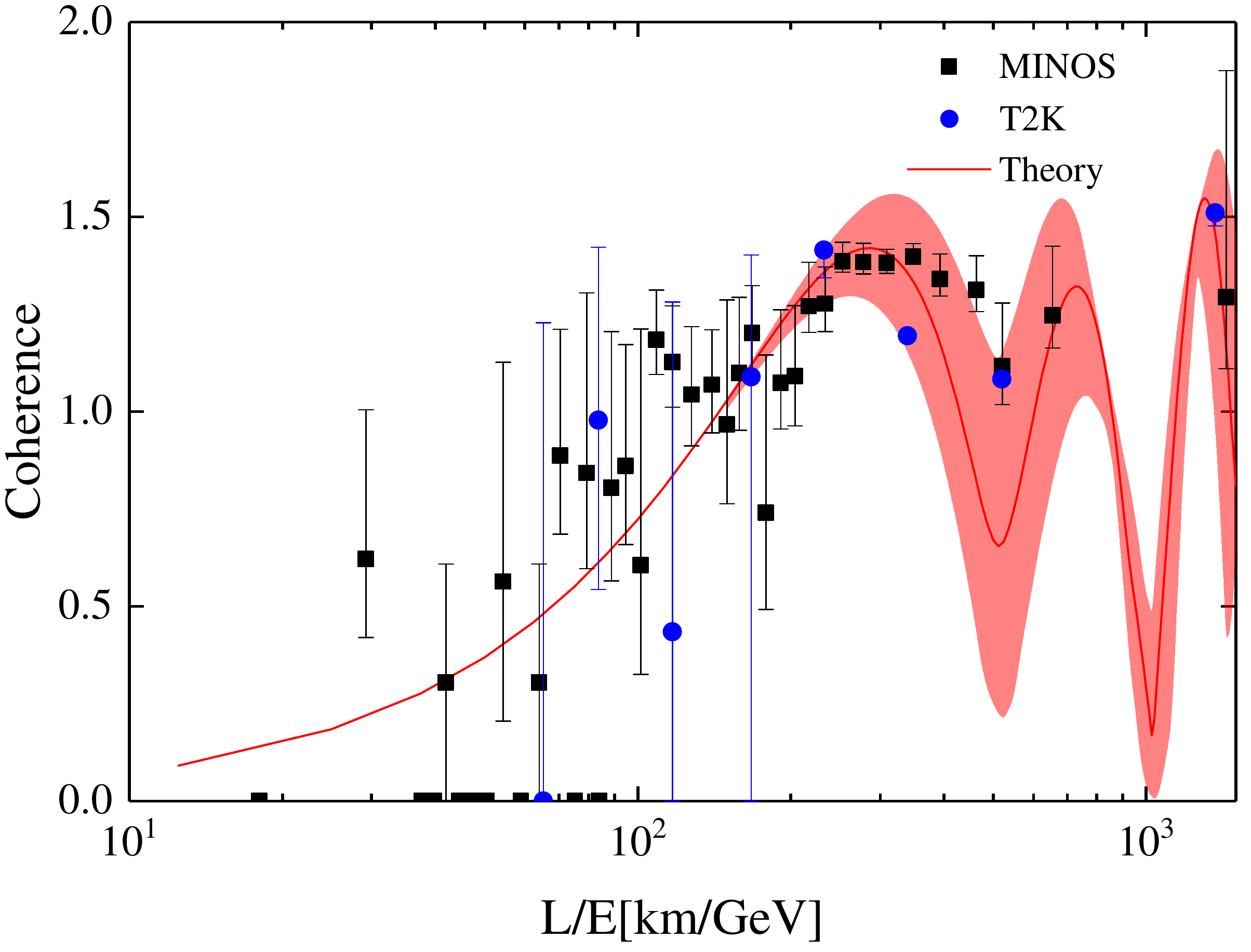}
\caption{ The coherence in theory (red, solid line) with a $3\sigma$ confidence interval around the
fitted prediction (red band), MINOS collaboration (black, square) and
$T2K$ collaboration (blue, circle) with error bar
for three-flavored neutrino oscillations as the function of ratio between
the distance and neutrino energy $L/E$ are shown. Taking the error bar into consideration, the data are in consistent with the theoretic $3\sigma$ range in the long-distance case, too.}\label{fig4}
\end{center}
\end{figure}

\emph{Coherence in muon antineutrino oscillations.---}
In the MINOS and T2K collaboration, muon (anti)neutrinos are produced from the proton beams of the accelerators. In the MINOS experiment, its baseline takes a longer distance of 735 $km$ and it
covers the energy from 0.5 to 50 $GeV$ \cite{minosdata}, which reveals that the ratio $L/E$ is in the range $[15, 1500]$. The T2K experiment demonstrates the oscillatory nature of neutrino flavor transformation by observing muon antineutrinos survival probability with energies of a few $GeV$
from nuclear reactors about 295 $km$ \cite{t2kdata}.

When the muon flavor state is prepared at initial time $t=0$,
the state of the time evolution for three-flavored neutrino oscillations is given by $|\nu_{\mu}(t)\rangle=a_{\mu e}(t)|\nu_{e}\rangle+a_{\mu\mu}(t)|\nu_{\mu}\rangle+a_{\mu\tau}(t)|\nu_{\tau}\rangle$.
The probabilities for finding the neutrino in state $|\nu_{e}\rangle$,
$|\nu_{\mu}\rangle$ and $|\nu_{\tau}\rangle$ are, respectively,
$P_{\mu e}(t)=|a_{\mu e}(t)|^2$, $P_{\mu\mu}(t)=|a_{\mu\mu}(t)|^2$ and $P_{\mu\tau}(t)=|a_{\mu\tau}(t)|^2$.
In Fig. \ref{fig3}, we show the survival  probabilities of the muon neutrino oscillation as a function of $L/E$. Similar to the case of electron neutrino, the coherence can be calculated as
\begin{align}  
\mathcal{C}_u=2\left(\sqrt{P_{\mu e}(t)P_{\mu\mu}(t)}+\sqrt{P_{\mu e}(t)P_{\mu\tau}(t)}+\sqrt{P_{\mu\mu}(t)P_{\mu\tau}(t)}\right),
\end{align}
Again, the coherence is determined primarily by the experimental data of the survival probability $P_{\mu\mu}$, theoretical values of the ratio $\zeta=P_{\mu\tau}(t)/P_{\mu e}(t)$ is employed. The coherence of the neutrino oscillation from the MINOS and T2K collaboration as a function of ratio $L/E$ is plotted in Fig. \ref{fig4}. The theoretic
coherence shows a more complicated behavior and the peaks appear at the points when the probabilities $P_{\mu \mu}(t)$ and $P_{\mu \tau}(t)$
are equivalent. It is nonvanishing in the neutrino propagation since the two of
the probabilities will be nonzero in spite of the trivial value of oscillation probability of electron neutrino $P_{\mu e}(t)$
compared to the survival probability $P_{\mu \mu}(t)$ and $P_{\mu \tau}(t)$. The experimental data for the coherence of MINOS
are distributed around the theoretical prediction line, which is consistent with the theoretical prediction of coherence.
On the other hand, the eight experimental data for coherence of the T2K collaboration also gives a good agreement with
theory.

\emph{Summary.---} In summary, we proposed a method for quantifying the quantumness of neutrino oscillations with the use of coherence measure developed in quantum resource theory. We compared the coherence in experimentally-observed neutrino oscillations from different sources, including Daya bay, Kamland, MINOS and T2K, which all exhibit good agreements with the theoretical predictions.  These results suggest that the coherence can be a reliable tool for the quantification of superposition in the three-flavored neutrino oscillation over a macroscopic distance of thousands of kilometers, certifying the quantumness of elementary particles other than photons.

\emph{Acknowledgements.---} This work is supported by Natural Science Foundation of Guangdong Province (2017B030308003) and the Guangdong Innovative and Entrepreneurial Research Team Pro- gram (No.2016ZT06D348), and the Science Technology and Innovation Commission of Shenzhen Municipality (ZDSYS20170303165926217, JCYJ20170412152620376), and the Postdoctoral Science Foundation
of China (No.2018M632195).


\begin{thebibliography}{}

\bibitem{Pontecorvo:1957qd}
  B. Pontecorvo,
  Sov.\ Phys.\ JETP {\bf 7}, 172 (1958)
  [Zh.\ Eksp.\ Teor.\ Fiz.\  {\bf 34}, 247 (1957)].

  \bibitem{Maki:1962mu}
  Z. Maki, M. Nakagawa and S. Sakata,
  Prog. Theor. Phys. \textbf{28}, 870 (1962).




  \bibitem{e1}
 Q. R. Ahmad \emph{et al.}  (SNO Collaboration),
  Phys. Rev. Lett. \textbf{89}, 011302 (2002).

\bibitem{e2}
M. Altmann \emph{et al.}  (GNO Collaboration),
  Phys. Lett. B \textbf{616}, 174 (2005).

\bibitem{e3}
S. Fukuda \emph{et al.} (Super-Kamiokande Collaboration),
  Phys. Lett. B \textbf{539}, 179 (2002).

 \bibitem{e4}
  T. Araki \emph{et al.}  (KamLAND Collaboration),
  Phys. Rev. Lett. \textbf{94}, 081801 (2005).


   \bibitem{e5}
  F. P. An \emph{et al. } (Daya Bay Collaboration),
  Phys. Rev. Lett.  \textbf{108}, 171803 (2012).

\bibitem{e6}
 M. H. Ahn \emph{et al.} (K2K Collaboration),
  Phys. Rev. D \textbf{74}, 072003 (2006).

\bibitem{e7}
P.~Adamson \emph{et al.} (MINOS Collaboration),
  Phys. Rev. Lett. \textbf{101}, 131802 (2008).

\bibitem{e8}
 P. Adamson \emph{et al.}  (MINOS Collaboration),
  Phys. Rev. Lett. \textbf{107}, 181802 (2011).



 \bibitem{e9}

  K. Abe \emph{et al.}  (T2K Collaboration),
  Phys. Rev. Lett. \textbf{107}, 041801 (2011).
\bibitem{Camilleri:2008zz}
  L. Camilleri, E. Lisi and J. F.Wilkerson,
  Ann. Rev. Nucl. Part. Sci. \textbf{58}, 343 (2008).


  \bibitem{Duan:2010bg}
  H. Duan, G. M. Fuller and Y. Z. Qian,
  Ann. Rev. Nucl. Part. Sci. \textbf{60}, 569 (2010).


\bibitem{dayabaydata} F. P. An, \emph{et al.} (Daya Bay Collaboration), Phys. Rev. Lett. \textbf{115}, 111802 (2015).

\bibitem{kamlanddata} A. Gando, \emph{et al.} (Kamland Collaboration), Phys. Rev. D \textbf{88}, 033001 (2013).

\bibitem{minosdata} A. B. Sousa (MINOS and MINOS+ Collaborations),  AIP Conf. Proc. \textbf{1666}, 110004 (2015).

\bibitem{t2kdata} K. Abe, \emph{ et al.} (The T2K Collaborations), Phys. Rev. D \textbf{96}, 011102(R) (2017).

\bibitem{Bustamante:2015waa}
  M. Bustamante, J. F. Beacom and W. Winter,
  Phys. Rev. Lett. \textbf{115}, 161302 (2015).

\bibitem{Huang:2015flc}
  Y. Huang and B. Q. Ma,
  The Universe \textbf{3}, 15 (2015).










\bibitem{Leggett} A. J. Leggett and A. Garg, Phys. Rev. Lett. \textbf{54}, 857 (1985).

\bibitem{Emary} C. Emary, N. Lambert and F. Nori, Rep. Prog. Phys. \textbf{77}, 039501 (2014).

\bibitem{Gangopadhyaytwo} D. Gangopadhyay, D. Home and A. S. Roy, Phys. Rev. A \textbf{88}, 022115 (2013).

\bibitem{Formaggio} J. A. Formaggio, D. I. Kaiser, M. M. Murskyj, and T. E. Weiss, Phys. Rev. Lett. \textbf{117}, 050402 (2016).

\bibitem{Fu} Q. Fu, X. Chen, Eur. Phys. J. C \textbf{77}, 775 (2017).

\bibitem{Gangopadhyaythree} D. Gangopadhyaya, A. S. Royb, Eur. Phys. J. C \textbf{77}, 260 (2017).

\bibitem{Baumgratz} T. Baumgratz, M. Cramer and M. B. Plenio, Phys. Rev. Lett. \textbf{113}, 140401 (2014).

\bibitem{Streltsov} A. Streltsov, G. Adesso and M. B. Plenio, Rev. Mod. Phys. \textbf{89}, 041003 (2017).

\bibitem{Hu} M. L. Hu, X. Hu, J. C. Wang, Y. Peng, Y. R. Zhang, and H. Fan, arXiv: 1703.01852 (2017).


\bibitem{winter} A. Winter and D. Yang, Phys. Rev. Lett. \textbf{116}, 120404 (2016).



\bibitem{matera} J. M. Matera, D. Egloff, N. Killoran, and M. B. Plenio, Quantum Sci. Technol. \textbf{1}, 01LT01 (2016).

\bibitem{piani} M. Piani, M. Cianciaruso, T. R. Bromley, C. Napoli, N. Johnston, and G. Adesso, Phys. Rev. A \textbf{93}, 042107 (2016).

\bibitem{gour} G. Gour, M. P. M\"{u}ller, V. Narasimhachar, R. W. Spekkens, N. Y. Halpern, Phys. Rep. \textbf{583}, 1 (2015).


\bibitem{Maki} Z. Maki, M. Nakagawa, S. Sakata, Prog. Theor. Phys. \textbf{28}, 870 (1962).

\bibitem{Garcia2} M. C. Gonzalez-Garcia, M. Maltoni and T. Schwetz,  J. High Energ. Phys. \textbf{1411}, 052 (2014).

\bibitem{Giunte} C. Giunte and C. W. Kim, Fundamentals of neutrino physics and astrophysics, Oxford University Press (2007).




\end{thebibliography}
\end{document}